# Effective Digital Image Watermarking in YCbCr Color Space Accompanied by Presenting a Novel Technique Using DWT

Mehdi Khalili[1] and David Asatryan[2]

Institute for Informatics and Automation Problems of NAS of RA
e-mail: Khalili@ipia.sci.am[1], dasat@ipia.sci.am[2]

**Abstract**

In this paper, a quantization based watermark casting and blind watermark retrieval algorithm operating in YCbCr color space using discrete wavelet transform (DWT), for ownership verification and image authentication applications is implemented. This method uses implicit visual masking by inserting watermark bits into only the wavelet coefficients of high magnitude, in Y channel of YCbCr color space. A blind watermark retrieval technique that can detect the embedded watermark without the help from the original uncorrupted image is devised which is computationally efficient. The new watermarking algorithm combines and adapts various aspects from existing watermarking methods. Experimental results show that the proposed technique to embed watermark provides extra imperceptibility and robustness against various signal processing attacks in comparison with the same technique in RGB color space.

## 1. Introduction

Watermarking means imperceptible embedding of information in an original multimedia data such as audio, video, text and image.

Nowadays, watermarking of images is becoming increasingly of interest in tasks such as copyright control, image identification, verification, and data hiding. One of the most popular which methods for image watermarking is Spread-spectrum watermarking where embeds a white noise watermark into transform coefficients of an image and verifies the presence of the watermark by measuring the correlation between the watermarked coefficients and the watermark sequence [1-5].

Perceptual invisibility, robustness against different image processing attacks, such as compression and geometric distortions, and ability of watermarking detection without ambiguity are three important requirements of a watermarking scheme [6].

In this paper, a novel quantization based digital image watermarking technique for ownership verification and image authentication applications is proposed. The motivation for this algorithm is based upon various aspects from watermarking schemes presented by [7, 8].

Dugad in [7] describes a watermarking technique that inserts the watermark in the most significant coefficients in the DWT domain and does not require the original image in the detection process. Dugad's technique uses an image sized watermark to negate the order dependence of significant coefficients in the detection process. However, it does not embed the watermark into many coefficients of the image, and also retrieval algorithm can tell only if the watermark is present or absent but it cannot recover the actual watermark.

Inoue in [8] describes a quantization based watermarking technique that inserts watermark into the detailed coefficients at the coarsest scales in the DWT domain and like the Dugad algorithm, the watermark is embedded in the perceptually significant coefficients, and requires a file to be saved detailing the locations where the watermark bits are embedded. In Inoue's technique the detector can recover the binary watermark sequence unlike Dugad algorithm, which can detect only the presence.

A variety of watermarking techniques has been proposed in recent years. The presented technique in this paper employs a watermark matrix equal to the size of the third level DWT matrix of the image in Y channel of YCbCr color space. The watermark insertion algorithm will overcome the disadvantage of additive insertion technique. Thus, for comparable robustness performance, this algorithm will produce watermarked images with less degradation that the Dugad technique and improves upon the Inoue technique by eliminating the position file in the recovery process.

Obtained results from experiments show the efficiency of presented technique in the field of digital images watermarking in wavelet transform domain. These results show that watermarking in YCbCr color space versus image processing attacks is more robust than watermarking in RGB color space, and has higher transparency [9].

## 2. Two Dimensional Wavelet Transform

The wavelet transform has been extensively studied in the last decade [10]. It is suitable for the watermarking applications because of its properties such as: (1) precise localization ability in space and in frequency [11], it means that the human perception splits and processes the image in several frequency channels as well as in multiresolution decomposition [12, 13], (2) excellent multi-scale analysis, and, (3) the publicly available masking thresholds [14].

For an input sequence of length n, DWT will generate an output sequence of length n [15]. In a 2-D wavelet transform, we can write the scaling function $\Phi_{LL}(x,y)$ of low-low subband as $\Phi_{LL}(x,y) = \Phi(x)\Psi(y)$, where $\Psi$ and $\Phi$ are the wavelet function and the scaling function respectively.

We also can obtain three other 2-dimensional wavelets by using the wavelet associated function $\Psi(x)$ as follows [15, 16, 17]:

$$\Psi_{LH}(x,y) = \Phi(x)\Psi(y); horizontal$$
$$\Psi_{HL}(x,y) = \Psi(x)\Phi(y); vertical$$
$$\Psi_{HH}(x,y) = \Psi(x)\Psi(y); diagonal$$

where H is a highpass filter and L is a lowpass filter.

In a DWT of 2-D signal (image), signal firstly is decomposed into the one approximation subband (LL1) and three details subbands (HL1, LH1, HH1) by cascading the signal horizontally and vertically with critically subsampled filter banks. Then, the approximation subband LL1 is decomposed again for obtaining the coarser-scaled wavelet coefficients. This process is repeated a number of arbitrary times which is determined by the application at hand. The signal can reconstruct from DWT coefficients. This reconstruction process is called the inverse DWT (IDWT).

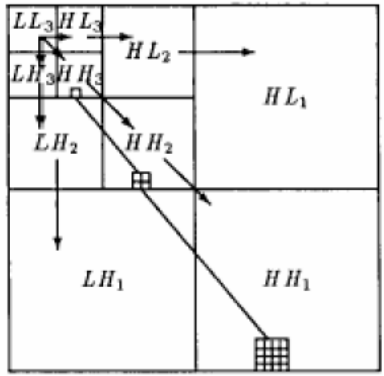

*Fig. 1. Three-level of a 2-D signal DWT decomposition with ten subbands.*

Fig. 1 shows a 2-D signal and its DWT decomposition. In this figure, signal is decomposed into three levels with ten subbands. Each level has various band-information such as low-low, low-high, high-low, and high-high frequency bands [15].
  Note that in Fig. 1, the lowest frequency subband is at the top left, and the highest frequency subband is at the bottom right.

## 3. A Brief Overview in YCbCr Color Space

YCbCr is a family of color spaces commonly used to represent digital in video and digital photography systems [18, 19]. The transformation simplicity and explicit separation of luminance and chrominance components make this color space attractive for skin color modeling. When a still image in RGB form is being prepared for encoding under the JPEG image encoding and compression system, it is transformed to the YCbCr color space. In this color space luminance (brightness) information is stored as a single component Y. Note that despite its symbol, Y does not represent the luminance of the color represented by R, G, and B, although it is something like it. It is sometimes called luma (a term drawn from television transmission technology).Chrominance (color) information corresponds to the two color-difference components Cb and Cr too [18, 19]. Cb is blue color minus luma (B−Y) and Cr is red color minus luma (R−Y) [20]. The basic equations to convert between RGB color space with YCbCr color space are well known (see, for example, [21]). In this color space, a pixel is classified as skin if the following conditions are satisfied [22]
$85 \leq Cb \leq 135, \quad 135 \leq Cr \leq 180, \quad Y \geq 80$.

## 4. Presented Watermarking Framework

The basic task of a digital watermarking is to make watermarks invisible to human eyes as well as robust to various attacks. The presented watermarking framework has a suitable performance to get this target. This framework is based on the discrete wavelet transform (DWT).
  The block diagram of the presented watermarking framework is shown in Fig 2. As it is seen, the host image is converted into YCbCr color space, and then the Y channel is decomposed into wavelet coefficients. Then, the watermark embedding process is performed in the approximation coefficients of the DWT of the host image.

## 4.1. Watermark Embedding Process

The algorithm for embedding watermark in the approximation coefficients of the host image Y channel is described as follows:

Step 1: Convert RGB channels of a host image W into YCbCr channels.

Step 2: Decompose the Y channel of converted host image into a three-level wavelet pyramid structure with ten DWT subbands, F(H).

Step 3: Save the signs of selection coefficients in a matrix sign.

Step 4: All the approximation coefficients with magnitude greater than $T_1$ and less than $T_2$ are selected, where $T_1$ and $T_2$ are the arbitrary threshold values.

Step 5: A binary watermark of the same size as the third wavelet level subband of the host image created using a secret key.

Step 6: Quantize absolute values of selection coefficients to embed the watermark. The value that the selected coefficients are quantized depends on then value of the watermark file whether it has 1 or 0 at that location. The selected wavelet coefficients, $w_{i,j}$, will be quantized to $T_1 + X_1$ if the value of the watermark mark file, $x_{i,j}$, is 0. Also the selected wavelet coefficients, $w_{i,j}$, will be quantized to $T_2 - X_1$ if the value of the watermark mark file, $x_{i,j}$, is 1. Thus adapting Inoue quantization method for watermark insertion:

$$\begin{cases} w_{i,j} = (w_{i,j})(T_1 + X_1) & \text{if } x_{i,j} = 0 \\ w_{i,j} = (w_{i,j})(T_2 - X_1) & \text{if } x_{i,j} = 1 \end{cases}.$$

Note that the $X_1$ is an arbitrary parameter, which narrows the range between the two thresholds $T_1$ and $T_2$ in order to aid robust oblivious extraction described in next section.

Step 7: Effect matrix sign into the embedded coefficients.

Step 8: Reconvert YCbCr channels of the changed host image into RGB channels.

Step 9: A watermarked image W' is then generated by inverse DWT with all changed and unchanged DWT coefficients.

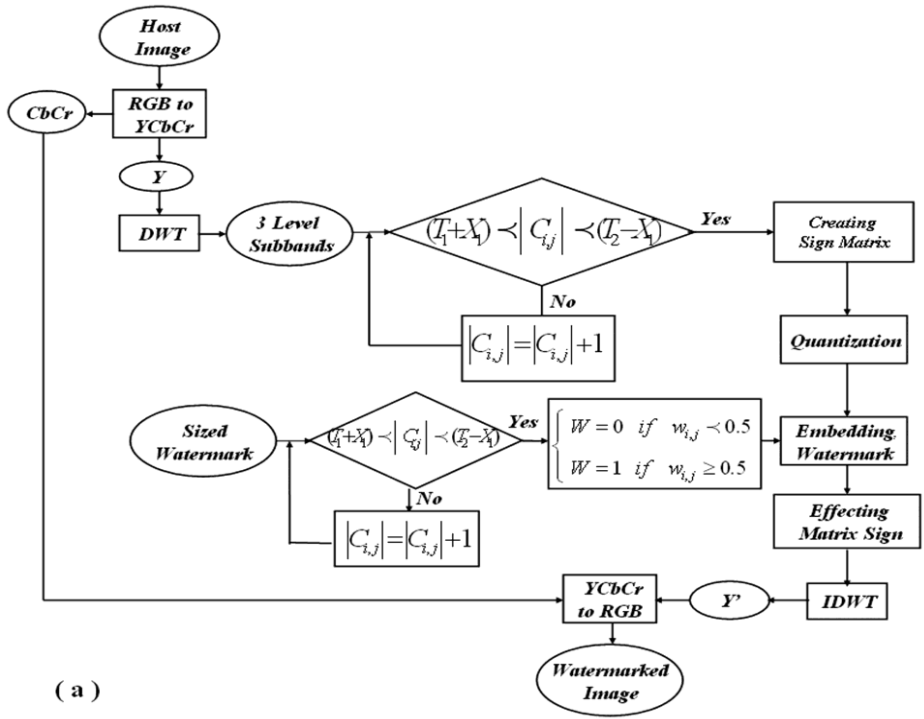

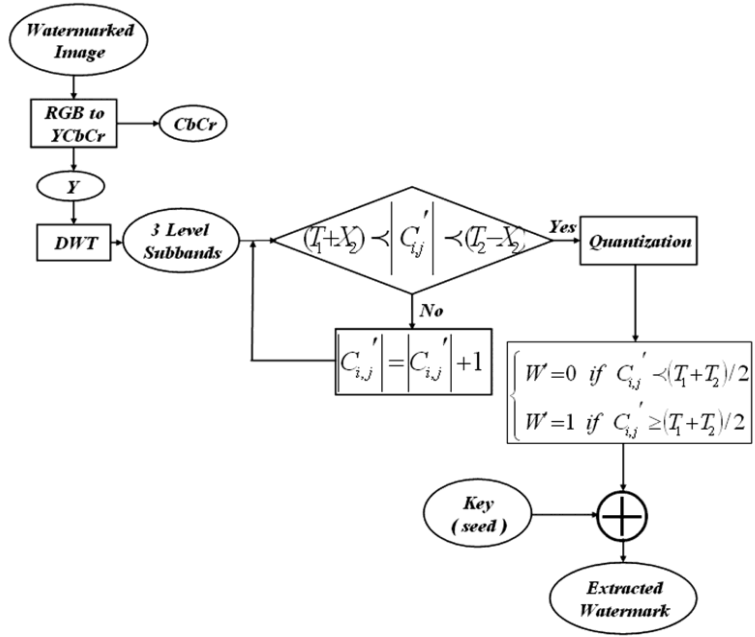

*Fig. 2. Block diagrams of the presented watermarking technique.*
*(a) Embedding process. (b) Extracting process.*

4.2. Watermark Extracting Process

The presented extracting process algorithm is blind. This algorithm is described as follows:
    Step 1: The RGB channels of the watermarked image are converted into YCbCr channels using the mentioned equations.
    Step 2: Decompose the Y channel of the converted watermarked image into a three-level DWT with ten subbands.
    Step 3: All the approximation coefficients in the third level of wavelet transform, of magnitude greater than or equal to $T_1 + T_2$ and less than or equal to $T_2 - X_2$ are selected. Here

$X_2$ is also an arbitrary parameter chosen to aid robust oblivious extraction and $X_2$ should always be less than $X_1$.

Step 4: Quantize absolute values of LL3 subband. In the embedding process, all the wavelet approximation coefficients magnitude greater than $T_1$ and less than $T_2$ are selected and then quantized to either $T_1 + X_1$ or $T_2 - X_1$. In the extracting process, all the wavelet approximation coefficients of magnitude greater than or equal to $T_1 + X_2$ and less than or equal to $T_2 - X_2$ are selected to be dequantized to extract the embedded watermark. Also, the unmarked coefficients are likely to drift into the range of selected approximation coefficients after an attack. The introduction of $X_1$ and $X_2$ parameters to the watermarking algorithm is very crucial since they give a degree of tolerance to the recovery algorithm to faithfully recover the embedded watermark.

Step 5: A watermark bit is decoded for each of the selected wavelet coefficients via the same process described by Inoue:

$$\begin{cases} 0 & \text{if} \quad w_{i,j} < (T_1 + T_2)/2 \\ 1 & \text{if} \quad w_{i,j} \geq (T_1 + T_2)/2 \end{cases}.$$

Step 6: Finally the recovered watermark is correlated with the original watermark file (obtained via secret key). This allows a confidence measure to be ascertained for the presence or absence of a watermark in an image.

## 5. Experimental Results

The presented watermarking framework was implemented for evaluating both properties of imperceptibility and robustness. Three 256×256 images: Lenna, Bust and, Peppers shown in Fig3 were taken as the host images to embed the binary watermark bits. For the entire test results in this paper, MATLAB R2007a software was used. Also for computing the wavelet transforms in the experiments, 7-9 point biorthogonal spline wavelet filters: {-0.0645, -0.0407, 0.4181, 0.7885, 0.4181, -0.0407, -0.0645} and {-0.0378, -0.0238, 0.1106, 0.3774, -0.8527, 0.3774, 0.1106, -0.0238, -0.0378} were used for wavelet decomposition and filters: {0.0378, -0.0238, -0.1106, 0.3774, 0.8527, 0.3774, -0.1106, -0.0238, 0.0378} and {-0.0645, 0.0407, 0.4181, -0.7885, 0.4181, 0.0407, -0.0645} were used for image reconstruction. By the way, in the experiments, the wavelet function bior4.4 was used as the mother wavelet function. Cause of use of B-spline function wavelet is that, B-spline functions, do not have compact support, but are orthogonal and have better smoothness properties than other wavelet functions [6]. Also, for the entire of experiments, $T_1$, $T_2$, $X_1$, $X_2$ were equal to 1500, 1600, 10 and 20, respectively.

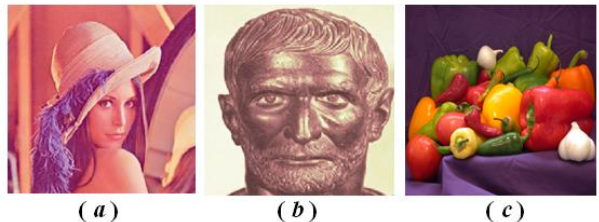

**Fig. 3.** The host images for watermarking. (a)-(c) Lenna, Bust, Peppers.

In the experiments, we measured the similarity of original host image x and watermarked images x' by the standard correlation coefficient (Corr) as [15]:
Moreover, the peak signal-to-noise ratio (PSNR) was used to evaluate the quality of the watermarked image. The PSNR is defined as [15, 23]:

$$PSNR = 10 \log_{10} \frac{255^2}{MSE} \quad (dB)$$

where mean-square error (MSE) is defined as [15, 23]:

$$MSE = \frac{1}{mn} \sum_{i=1}^{m} \sum_{j=1}^{n} (h_{i,j} - h'_{i,j})^2.$$

where $h_{i,j}$ and $h'_{i,j}$ are the gray levels of pixels in the host and watermarked images, respectively.

The larger PSNR is, the better the image quality is. In general, a watermarked image is acceptable by human perception if its PSNR is greater than 30 dB. In other words, the correlation is used for evaluating the robustness of watermarking technique and the PSNR is used for evaluating the transparency of watermarking technique [15].

We also used the normalized correlation (NC) coefficient to measure the similarity between original watermarks W and the extracted watermarks W' that is defined as [15, 24]:

$$NC = \frac{\sum_i \sum_j w_{i,j} * w'_{i,j}}{\sum_i \sum_j w_{i,j}^2}.$$

The proposed watermarking approach yields satisfactory results in watermark imperceptibility and robustness. Table 1 shows the extracting results and the watermarked images using the proposed method without any attacks. As it is visible the watermarked results are excellent. The PSNR of the watermarked images produced by the presented technique are all greater than 47 dB, NC between original watermark images and extracted watermark images are all equal 1, the error rates are all 0, and correlations between host images and watermarked images are all greater than 0.999, which are perceptually imperceptible as shown in Fig 4.

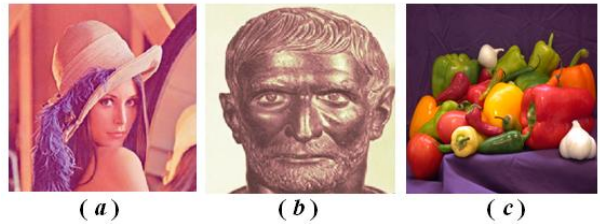

**Fig. 4.** The watermarked images produced by the presented technique.

### 5.1. Robustness Examinations

After getting to the desired fidelity, robustness of the watermark against JPEG-compression and signal processing attacks were tested.

#### 5.1.1. *Robustness against JPEG-compression attacks*

Table 2 shows the extracted results from JPEG-compressed version of watermarked images with compression ratios (CR) 20, 15, 10 and 5. The quality of watermarked images is still in good situation even under the high compression ratio. In the proposed technique, the error rate of the extracted watermarks is rarely small and Normalized Correlation is still suitable, even under the situation of compression ratio 15.

#### 5.1.2. *Robustness to geometrical rotation attack*

After testing robustness against JPEG-compression attacks, robustness of watermarked image was examined by rotation attack. The image was attacked by geometrical rotation by different angles. Table 3 shows the extracted results from attacks. As it is seen, the error rate and Normalized Correlation are still suitable in different angles.

**Table 1.** Obtained results of watermark extracting.

| Image | Lenna | Bust | Peppers |
|---|---|---|---|
| Number of Embedded Watermark Bits | 151 | 126 | 77 |
| Correlation | 0.9999 | 0.9999 | 0.9999 |
| PSNR (dB) | 47.0542 | 51.1180 | 63.9543 |
| Number of Extracted Watermark Bits | 151 | 126 | 77 |
| Normalized Correlation | 1.00 | 1.00 | 1.00 |
| % Error Rate | 0 | 0 | 0 |

**Table 2.** Obtained results of JPEG-compression attacks.

| Image | CR | Number of Extracted watermark Bits | Normalized Correlation | % Error Rate |
|---|---|---|---|---|
| Lenna | 5 | 97 | 0.989 | 35.76 |
|  | 10 | 102 | 0.991 | 32.45 |
|  | 15 | 139 | 0.997 | 7.94 |
|  | 20 | 144 | 0.998 | 4.63 |
| Bust | 5 | 73 | 0.981 | 42.06 |
|  | 10 | 96 | 0.993 | 23.80 |
|  | 15 | 111 | 0.996 | 11.90 |
|  | 20 | 121 | 0.998 | 3.96 |
| Peppers | 5 | 44 | 0.980 | 42.85 |
|  | 10 | 53 | 0.990 | 31.16 |
|  | 15 | 69 | 0.996 | 10.38 |
|  | 20 | 72 | 0.997 | 6.49 |

**Table 3.** Obtained results of Geometrical Rotation attacks.

| Image | | | | | |
|---|---|---|---|---|---|
| Lenna | Angle (Degrees) | 20 | - 60 | 130 | - 110 |
|  | Number of Extracted watermark Bits | 133 | 109 | 128 | 122 |
|  | Normalized Correlation | 0.996 | 0.994 | 0.994 | 0.992 |
|  | % Error Rate | 11.92 | 27.81 | 15.23 | 19.20 |
| Bust | Angle (Degrees) | 20 | - 60 | 130 | - 110 |
|  | Number of Extracted watermark Bits | 107 | 88 | 101 | 95 |
|  | Normalized Correlation | 0.994 | 0.990 | 0.993 | 0.992 |
|  | % Error Rate | 15.0 | 30.15 | 19.84 | 24.60 |
| Peppers | Angle (Degrees) | 20 | - 60 | 130 | - 110 |
|  | Number of Extracted watermark Bits | 69 | 53 | 62 | 59 |
|  | Normalized Correlation | 0.996 | 0.990 | 0.993 | 0.992 |
|  | % Error Rate | 10.38 | 31.16 | 19.48 | 23.37 |

### 5.1.3. *Robustness to filtering*

Here, robustness of watermarked images was examined with median filter. The median filter is generally used to eliminate noises. Table 4 shows the results of the extracted watermarks under median filter testing. From this table it is visible that, the presented technique can satisfy watermarking targets.

**Table 4.** Obtained results of JPEG-compression attacks.

| Image | Matrix | Number of Extracted watermark Bits | Normalized Correlation | % Error Rate |
|---|---|---|---|---|
| Lenna | 3×3 | 150 | 0.9959 | 0.66 |
| | 5×5 | 149 | 0.9948 | 1.32 |
| Bust | 3×3 | 124 | 0.9369 | 1.58 |
| | 5×5 | 122 | 0.9359 | 3.17 |
| Peppers | 3×3 | 76 | 0.9999 | 1.29 |
| | 5×5 | 75 | 0.9998 | 2.59 |

**Table 5.** Obtained results of watermark extracting in RGB color space.

| Image | Lenna | Bust | Peppers |
|---|---|---|---|
| Number of Embedded Watermark Bits | 48 | 52 | 8 |
| Correlation | 0.9997 | 0.9999 | 0.9999 |
| PSNR (dB) | 44.1349 | 46.6938 | 58.2391 |
| Number of Extracted Watermark Bits | 48 | 52 | 8 |
| Normalized Correlation | 1.00 | 1.00 | 1.00 |
| % Error Rate | 0 | 0 | 0 |

**Table 6.** Obtained results of JPEG-compression attacks in RGB color space.

| Image | CR | Number of Extracted watermark Bits | Normalized Correlation | % Error Rate |
|---|---|---|---|---|
| Lenna | 5 | 14 | 0.914 | 70.83 |
| | 10 | 21 | 0.951 | 56.25 |
| | 15 | 24 | 0.957 | 50.00 |
| | 20 | 27 | 0.961 | 43.75 |
| Bust | 5 | 18 | 0.921 | 65.38 |
| | 10 | 26 | 0.962 | 50.00 |
| | 15 | 29 | 0.963 | 44.23 |
| | 20 | 32 | 0.969 | 38.46 |
| Peppers | 5 | 2 | 0.982 | 75.00 |
| | 10 | 4 | 0.994 | 50.00 |
| | 15 | 5 | 0.997 | 37.50 |
| | 20 | 5 | 0.997 | 37.50 |

**Table 7.** Obtained results of Geometrical Rotation attacks in RGB color space.

|  | Angle (Degrees) | 20 | - 60 | - 110 | 130 |
|---|---|---|---|---|---|
| Lenna | Number of Extracted watermark Bits | 31 | 29 | 30 | 29 |
|  | Normalized Correlation | 0.963 | 0.943 | 0.956 | 0.943 |
|  | % Error Rate | 35.41 | 39.58 | 37.50 | 39.58 |
| Bust | Angle (Degrees) | 20 | - 60 | - 110 | 130 |
|  | Number of Extracted watermark Bits | 39 | 28 | 33 | 31 |
|  | Normalized Correlation | 0.952 | 0.922 | 0.946 | 0.937 |
|  | % Error Rate | 25.00 | 46.15 | 36.53 | 40.38 |
| Peppers | Angle (Degrees) | 20 | - 60 | - 110 | 130 |
|  | Number of Extracted watermark Bits | 2 | 2 | 2 | 1 |
|  | Normalized Correlation | 0.907 | 0.907 | 0.907 | 0.868 |
|  | % Error Rate | 75.00 | 75.00 | 75.00 | 87.50 |

**Table 8.** Obtained results of Geometrical Rotation attacks in RGB color space.

|  | Matrix | Number of Extracted watermark Bits | Normalized Correlation | % Error Rate |
|---|---|---|---|---|
| Lenna | 3×3 | 43 | 0.988 | 10.41 |
|  | 5×5 | 35 | 0.983 | 27.08 |
| Bust | 3×3 | 42 | 0.935 | 19.23 |
|  | 5×5 | 43 | 0.936 | 17.30 |
| Peppers | 3×3 | 7 | 0.999 | 12.5 |
|  | 5×5 | 6 | 0.998 | 25.00 |

Comparing the previous results and Tables 6, 7 and 8, it is found that the presented framework is obviously more transparent and more robust than the usual frameworks.

As it is visible from table 5, payload is decreased and transparency becomes slightly severe because of its PSNR. Moreover in detection process, error rate is very great. It means that, the numbers of correct extracted bits is less.

## 6. Performing Presented Technique in RGB Color Space

To consider the presented technique efficiency, the watermarking process was performed in RGB color space by the same method, host images, and values that in YCbCr color space, and obtained results were compared with previous results. The obtained results of watermark extraction and robustness examinations are shown in tables 5, 6, 7 and 8.

## 7. Conclusions

In this paper, a novel digital image watermarking technique was presented where different host images of size 256X256 were watermarked in YCbCr color space using DWT, and watermark was extracted after signal processing attacks. A watermarking technique, which uses the concepts from Dugad technique and Inoue technique, was successfully implemented. Increase of the payload of inserted watermark, extra imperceptibility, and robustness against various signal processing attacks in comparison with the same technique in RGB color space were achieved. Summarization of the watermarking steps are as follows: first the host image is converted into YCbCr color space and then Y channel of converted image transformed in the wavelet domain using 7-9 biorthogonal spline filter coefficients. Then the transformed image was watermarked using the presented technique. Later the watermarked image was subjected to various signal processing attacks like JPEG-compression, filtering, and rotation. Then inserted watermark was retrieved from the attacked image. All the above listed steps were successfully implemented. The choice of $T_1$ and $T_2$ is very crucial; an appropriate choice of $T_2$ will be 83% to 87% of the maximum coefficient value and $T_1$ will be 43% to 47% of the maximum coefficient value.

Furthermore, the presented watermarking technique was performed by traditional methods (in RGB color space) with the same presented technique. Comparing the obtained results show that, the presented technique in YCbCr color space satisfy achievement to the watermarking targets.

## Գունավոր պատկերի՝ YCbCr տարածության մեջ DWT կիրառմամբ թվային ջրանշման նոր արդյունավետ մեթոդ

Մ. Խալիլի և Դ. Ասատրյան

### Ամփոփում

Հոդվածում առաջարկվել է չարտոնված օգտագործումից թվայնացված պատկերի պաշտպանության ալգորիթմ՝ երկչափանի ընդհատ վեյվլետ-ձևափոխությունների (DWT) կիրառմամբ։ Ալգորիթմը մշակվել է գունավոր պատկերի YCbCr տարածության համար, ընդ որում ջրանիշի բիտը ներդրվում է այդ տարածության Y բաղադրիչի առավել մեծ մագնիտուդով վեյվլետ-գործակիցների մեջ։ Ցույց է տրվել, որ առաջարկված մեթոդը ապահովում է ջրանիշի ավելի բարձր անտեսանելիություն և կայունություն տարբեր տեսակի հարձակումների նկատմամբ, քան գունավոր պատկերի RGB տարածության մեջ գործող նույնատիպ ալգորիթմները։